\documentclass[11pt]{amsart}
\usepackage[utf8]{inputenc}
\usepackage{amsfonts}
\usepackage{amsmath}
\usepackage{amssymb}
\usepackage{mathrsfs} 
\usepackage{enumerate}
\usepackage{xcolor}
\usepackage{rotating}
\usepackage{hyperref}

\usepackage{scalerel,stackengine}
\stackMath
\newcommand\rwhat[1]{%
\savestack{\tmpbox}{\stretchto{%
  \scaleto{%
    \scalerel*[\widthof{\ensuremath{#1}}]{\kern-.7pt\bigwedge\kern-.7pt}%
    {\rule[-\textheight/2]{1ex}{\textheight}}%WIDTH-LIMITED BIG WEDGE
  }{\textheight}% 
}{0.5ex}}%
\stackon[1pt]{#1}{\tmpbox}%
}
\usepackage{stackengine}
\newcommand\xrowht[2][0]{\addstackgap[.5\dimexpr#2\relax]{\vphantom{#1}}}

\usepackage{tikz}
\usetikzlibrary{arrows.meta,automata,quotes}

\theoremstyle{plain}
\newtheorem{thm}{Theorem}%[section]

\newtheorem{lem}[thm]{Lemma}

\theoremstyle{definition}
\newtheorem{defi}[thm]{Definition}

\newtheorem{example}[thm]{Example}

\newtheorem{obs}[thm]{Observation}

\newcommand{\tb}{\ensuremath{\mathrm{TB}}}

\renewcommand{\L}{\mathrm L}
\newcommand{\R}{\mathrm R}

\newcommand{\nL}{\ensuremath{\mathscr L}}
\newcommand{\nR}{\ensuremath{\mathscr R}}
\newcommand{\nN}{\ensuremath{\mathscr N}}
\newcommand{\nP}{\ensuremath{\mathscr P}}
\newcommand{\nl}{\ensuremath{\mathscr L}\;}
\newcommand{\nr}{\ensuremath{\mathscr R}\;}

\newcommand{\GL}{G^{\mathcal L}}
\newcommand{\GR}{G^{\mathcal R}}
\newcommand{\HL}{H^{\mathcal L}}
\newcommand{\HR}{H^{\mathcal R}}

\newcommand{\cg}[2]{\left\{ #1\!\mid \!#2\right\}}
\newcommand{\up}{\ensuremath{\uparrow}}
\newcommand{\down}{\ensuremath{\downarrow}}
\renewcommand{\emptyset}{\varnothing}
\newcommand{\nil}{\varnothing}
\renewcommand{\hat}{\widehat}
\renewcommand{\tilde}{\widetilde}

\newcommand{\B}{\mathcal B}
\renewcommand{\ge}{\geqslant}
\renewcommand{\le}{\leqslant}

\title{Bidding combinatorial games}

\author{Prem Kant}
\address{Prem Kant, IIT Bombay, India}
\email{premkant@iitb.ac.in}
\author{Urban Larsson}
\address{Urban Larsson, IIT Bombay, India}
\email{larsson@iitb.ac.in}
\author{Ravi K. Rai}
\address{Ravi Kant Rai, University of Liverpool, UK}
\email{r.k.rai@liverpool.ac.uk}
\author{Akshay V. Upasany}
\address{Akshay Vilas Upasany, IIT Bombay, India}
\email{akshay.upasany@iitb.ac.in}

\date{16 July 2022}

\begin{document}

\begin{abstract}
Combinatorial Game Theory is a branch of mathematics and theoretical computer science that studies sequential 2-player games with perfect information. Normal play is the convention where a player who cannot move loses. Here, we generalize the classical alternating normal play to infinitely many game families, by means of discrete Richman auctions (Develin et al. 2010, Larsson et al. 2021, Lazarus et al. 1996). We generalize the notion of a perfect play outcome, and find an exact characterization of outcome feasibility. As a main result, we prove existence of a game form for each such outcome class; then we describe their lattice structures. By imposing restrictions to the general families, such as impartial and {\em symmetric termination}, we find surprising analogies with alternating play.  
\end{abstract}

\maketitle
\section{Introduction}
Normal play combinatorial games are sequential games, usually played under the alternating move convention, and where ``the player who moves last wins'' \cite{BCG2001, C1976, S2013}. Or wait, should we rather say, ``a player who cannot move loses''? Of course, in alternating play, these statements are equivalent playing from a non-terminal position. The former is more intuitive from a recreational play perspective, but the latter is the correct one, from a recursive point of view;  the neutral element, the 0-game, has no move options for either player and so the current player loses. Moreover, if $0$ is the starting position, there is no ``last player''. The recursive point of view is fundamental to the theory, in computing perfect play outcomes via backward induction. 

We generalize alternating play Combinatorial Game Theory (CGT) into an infinite class of game families under the normal play convention, by combining the standard game forms with a certain bidding convention, the so-called {\em Richman auction} \cite{LLPU1996}. As usual, the players are called Left (female) and Right (male).
%, and \done{ the game will end in a finite number of moves.}
If a game $G$ is played by the discrete bidding convention \cite{DP2010,LPR2021}, then there is a total budget $\tb\in\mathbb N_0=\{0,1,\ldots\}$ partitioned between the players as $(p, q)$, with $p+q=\tb$, and where Left's part of the budget is $p$. The player who bids more moves next, and if bids are equal then a tiebreaker determines {\em the current player} (as in standard CGT we identify `next' with `current'). 

At each stage of play, exactly one of the players holds a tiebreaker, a tiebreaking {\em marker}, and this is denoted by $(G,\hat p)$ (Left holds the marker in the game $G$) or $(G,p)$ (if Right holds the marker). If a player wins a bid strictly, then this player hands over the bidding amount to the other player and plays their move. If bids are equal then the player with the marker wins the bid, they play their move, and the marker together with the bid shifts over to the other player. 

The marker holder can also choose to explicitly announce the marker with the bid. In this case, if the marker holder  wins the bid, they play their move, and the marker together with the bid shifts to their opponent. This additional rule will be important in restoring some familiar normal play properties. To illustrate this, consider the game $*=\cg{0}{0}$. In this game both Left and Right have only one option, which is to end the game.\footnote{In this paper we use some of the well known symbols from alternating normal play, such as $*,\up$ and so on. Here they refer to literal form games and never `game values'.} This game  should be favourable to the currently `stronger' 
player (first player in alternating play). 
Without this rule, since no player wants the tiebreaker if they have no move, the marker holder would benefit from a tie, and the other player would benefit from a strict win or loss of the bid.  This situation is a normal form (matrix) game with no pure equilibrium. Thus we cannot use any familiar tool from normal play theory. We will also see that this additional rule establishes that ``last move wins'' is the same as saying  ``cannot move loses''. Examples~\ref{ex:mix} and~\ref{ex:pure} in Section~\ref{sec:firstexamples} elaborate on this theme. 

The main results of this paper state that, given any total budget, a perfect play outcome must satisfy a defined {\em feasibility} (Section~\ref{sec:out}). And each feasible outcome class appears as the outcome of a bidding game (Section~\ref{sec:outappear}). 
 
\section{The basic set up}\label{sec:basicsetup}
 Given a total budget \tb, let us define $\B = \{0,\ldots, \tb,\hat 0,\ldots,  \hat \tb\}$, the set of all feasible player budgets. Here the word ``budget" includes the information of the marker holder. A game is a triple $(\tb, G, \tilde p)$, where we take a note of Left's part of the budget, $\tilde p\in \B $.  If $\tb$ is understood, we write $(G, \tilde p)$.  If we do not wish to specify $\tilde p$, we may write $(\tb, G)$ or just $G$. 
Game forms are recursively defined, with $G=\cg{\GL}{\GR}$, where $\GL$ is the set of all Left options and $\GR$ is the set of all Right options. If $\GL=\varnothing$ or $\GR=\varnothing$ then $G$ is Left terminal or Right terminal respectively, i.e., the current player, Left or Right, cannot move. In the case when  $\GL=\GR= \varnothing$, then $G=0$ is terminal irrespective of move order. A typical Left (Right) option of a game form $G$  is written  $G^L$ ($G^R$). In a play situation, we include Left's part of the budget in the notion of an option. Games are finite and contain no cycles. That is, each game has finitely many options, and the birthday (rank of game tree) is finite, implying that each play sequence is finite; this is also known as `short games' \cite{S2013}. The game $H$ is a \emph{subgame/follower}  of a game $G$ if there exists a path of moves, perhaps empty, (in any order of play) from $G$ to $H$.
 
A player obviously does not want to win a bid if they cannot move. And if they do not hold the marker they must {\em pass} in any such situation, that is, they must bid zero.
% In such situations, if the player without the marker bids $0$, we refer to this as a \emph{pass}. In particular nobody wants to win the Richman auction from a terminal position.

Consider a non-terminal position $(G,\hat p)$ (Left holds the marker). There are four cases to establish the next position:
\begin{itemize}
\item $(G,\hat p)\rightarrow (G^L,\rwhat{ p-\ell})$;  a Left move after bidding more, i.e. $\ell>r$.
\item $(G, \hat p)\rightarrow (G^L, p-\ell)$;  a Left move after bidding more, i.e. $\ell>r$, and including the marker. 
\item $(G, \hat p)\rightarrow (G^L, p-\ell)$;  a Left move after winning a tie, i.e. $\ell = r$. 
\item $(G, \hat p)\rightarrow (G^R, \rwhat{ p+r})$;  a Right move after bidding more, i.e.  $\ell<r$.
\end{itemize}

Observe that in case of a tie, the marker is transferred. This automatic rule is at the core of our generalization of alternating play. Namely $\tb=0$ corresponds to alternating normal play rules.

%Let us prove that bidding games generalize alternating play.

\begin{thm}\label{thm:tb0}
Consider $\tb = 0$. Then bidding play is identical to alternating play. The current player is the player who holds the marker.
\end{thm}
\begin{proof}
By the rules of bidding play, the marker holder shifts if the players tie a bid. Since $\tb=0$, every bid is a tie and the marker holder is the current player. 
\end{proof}

In the classical continuous Richman auctions \cite{LLPU1996}, the tiebreaker has infinitesimal value with respect to any bid, and so the theory is not sensitive to variations on tiebreaker transfer. We adopt a discrete bidding convention, which is more natural from a recreational play point of view, but where the analysis of the tiebreaker becomes non-trivial. In fact, the normal play winning implication in the first paragraph of this paper does not a priori hold, and moreover ambiguity on a winning strategy may appear, so traditional requirements for combinatorial games such as pure strategy subgame perfect equilibria may break. With some care of picking a sound tiebreaking convention this issue will be remedied, so that standard backward induction techniques (Section~\ref{sec:FFT}) will produce optimal outcomes that generalize normal play outcomes with new lattice structures (Section~\ref{sec:lattice}). 

\section{Games in pure subgame perfect equilibria}\label{sec:firstexamples}

Let us review why the tiebreaking rule used in \cite{LPR2021} would require mixed strategies in equilibrium in this study. A basic observation is that, independently of specifics of rule, the $0$-game is losing for the player who holds the tiebreaking marker. Namely, the player without the marker will bid 0, in a successful attempt to lose that bid. And indeed, this resembles alternating normal play, where the set of 0-games is the unique equivalence class of zugzwang positions, where no player has incentive to move. 

Let us dwell a bit more on the choice of tiebreaker. Two motivating examples will guide us further up the road. 
\begin{example}\label{ex:mix}
Consider $\tb=2$ and the game $*=\cg{0}{0}$ under the Richman bidding convention where the tiebreaker marker shifts player when used to resolve a tie, but not otherwise. Suppose  Left has budget \$1 together with the marker. Both
players have the terminal game as their only option, and so both players prefer not to
%here terminal game being the only option is important
own the marker at the next position. This means that Left wants a tie, but Right wants either player to win the bid strictly. There are two possibilities for Left to reach her goal, and there are two ways for Right. Namely Left prefers the bidding pair $(1,1)$ or $(0,0)$, whereas Right initially wants $(1,0)$ or $(0,1)$. See Table~\ref{tab:mix}. This situation violates the normal play assertion: ``last move wins" in case Left wins the bid by the bidding pair $(1,0)$. She gets the last move but loses the game. 

\begin{table}[ht]
\center{\caption{Consider the bidding convention in Example~\ref{ex:mix}. The table displays the winner of the game $*=\cg{0}{0}$ on total budget $\tb=2$, where the players own \$1 each, and where Left holds the tiebreaking marker. Rows and columns corresponds to the initial bids of Left and Right player respectively. Left plays rows and Right plays columns.}\label{tab:mix}
\begin{tabular}{|c|c|c|}
\hline % perhaps we want to use the form \bud{1}{4}
 Bids & $0$ & $1$   \\ \hline\hline
 $0$ & L & R  \\ \hline
 $1$  & R & L   \\ \hline
\end{tabular}}
\end{table}
\end{example}

Moreover, this bidding rule gives too much emphasis on the marker; both players' goal of the game would be to not hold the marker when the game ends, rather than focusing on ``who moves last". But the marker was introduced merely as a device to resolve ties. Hence, in this paper we will adapt the bidding convention in Example~\ref{ex:pure}, where Left can assure a win of the game $*$ by going all in at the second last bid; the last move wins.

\begin{example}\label{ex:pure}
We alter the bidding convention in Example~\ref{ex:mix}, so that the marker holder, here Left, may include the marker in her bid, and hand it over in case of winning the bid strictly. In case of a tie, of course, the marker swap is still mandatory. 
  With total budget $\tb=2$, Left has four bidding alternatives to start with, as displayed in Table~\ref{tab:pure}. The final winner is displayed in each case, and depending on Right bidding \$0 or \$1. Note that row 4 is better for Left than all other rows, regardless of Right's choice of move, so Left will start by bidding  $\widehat 1$. She gets the last move and wins the game.

\begin{table}[ht]
\center{\caption{The table displays the winner using the bidding convention in Example~\ref{ex:pure}, and otherwise the game configuration  as in Table~\ref{tab:mix}.}\label{tab:pure}
\begin{tabular}{|c|c|c|}
\hline \xrowht{9pt}
 Bids & $0$ & $1$   \\ \hline\hline\xrowht{9pt}
 $0$ & L & R  \\ \hline\xrowht{9pt}
 $1$  & R & L   \\ \hline\xrowht{9pt}
 $\widehat 0$  & L & R \\ \hline\xrowht{9pt}
 $\widehat 1$  & L & L  \\ \hline
\end{tabular}}
\end{table}
\end{example}

\section{The first fundamental theorem of bidding play}\label{sec:FFT}
Let us mention a couple of possibilities for discrete bidding and its tie-breaking marker. 

\begin{enumerate}[(i)]
    \item[(a)] a player who wins the bid decides who is the current player;
    \item[(b)] a player who wins the bid is the current player;
    \item the marker alternates between the players;
    \item the marker holder may include it in the bid;
    \item the marker stays with one of the players;
    \item the marker holder shifts if and only if it has been used to resolve a tie.
    \item if the marker holder wins the bid, the marker gets transferred. 
\end{enumerate}

In \cite{DP2010}, they study the variation (a), and in \cite{LPR2021},  (b) is combined with (iv).

Let us begin by proving that, for every game $G$, our variation  (b) together with (ii) has pure strategy subgame perfect bidding equilibria. 

\begin{thm}[First Fundamental Theorem]\label{thm:pure}
Consider the bidding convention where the tie-breaking marker may be included in a bid. For any game $(\tb, G,\tilde p)$ there is a pure strategy subgame perfect equilibrium (PSPE), computed by standard backward induction.
\end{thm}
\begin{proof}
 A terminal game $G=\cg{\nil}{\nil}$ has a pure strategy equilibrium; a player who holds the marker but does not have any option loses when the other player bids $0$. 

Suppose that $(G,\tilde p)$ is non-terminal. By induction, we may assume that each option is in a pure strategy subgame perfect equilibrium.  

There are four possibilities from the current position of the game $(G,\tilde p)$ :
\begin{enumerate}[i)]
    \item Left wins the current bid and reaches an option from where she wins the game using induction.
    \item Left is forced to win the current bid and from all her option she loses the game by induction.
    \item Left loses the current bid but from each Right options, she wins the game using induction.
    \item Left is forced to lose the current bid and Right plays to an option from where Left loses the game by induction.
\end{enumerate}
These four possibilities can be analysed by considering the following two cases:
\begin{enumerate}[{Case} a):]
    \item Suppose that the winner of the bid loses the game. Note that this case will arise only when a player is forced to win the bid because they hold the marker. In this case the other player has no incentive to change their bid (which is 0) and the one who is winning the bid cannot avoid winning the bid by changing their bid.
    \item Suppose that the winner of the bid wins the game.  
    Without loss of generality, say Left wins $(G,\tilde p\,)$, by bidding $\tilde\ell$. Clearly Left does not have any incentive to change her bid. By the assumption of this case, Right does not win the game if he outbids Left or ties in case of holding the marker. So, the remaining possibilities of a Right deviation is that he could lower his bid or tie in case of Left holding the marker. 

Case b1: Left holds the marker. Right's assumed bid is $r\le\ell $. If Right ties Left's bid, i.e. $r=\ell$, then Left wins the bid by marker holdership. Right can deviate by lowering his bid. But this would not change the outcome, since, if it would, Left would have included the marker in a strict winning bid. In case $r<\ell$, Left remains the winner of the bid. 

Case b2: Left does not hold the marker. Then the winning of the bid is strict. If Right lower his bid then this will not change the winner of the bid, and he will keep the marker.
\end{enumerate}
Since no player has incentive to change the bid, we have proved that $(G, \tilde p)$ has a PSPE. And the correctness of the backward induction approach follows by the method of proof.
\end{proof}

From now on, we will study the variation (b) together with (ii). 

By this result, henceforth we will refer to {\em perfect play} with the meaning perfect pure bids and a perfect move by the player who wins the bid.

Bids are typically simultaneous, but in view of perfect play and Theorem~\ref{thm:pure}, the order of bidding is irrelevant.

Consider a given stage of play.  A player who has the larger part of the budget, or half the budget together with the marker, is called the (currently) {\em dominating player}. 
%A player is {\em dominating} if they have the larger budget, or in case of holding the marker, the weakly larger budget. 
A player is {\em strictly dominating} if they have a strictly larger budget than their opponent.

Now we establish that ``last move wins'' is the same as ``cannot move loses".
%\pr{The issue with this lemma is, we are trying to establish last move wins is same as cannot move loses, when there is optimal play. But we should't just do this for optimal play. }
\begin{lem}[Last Move Wins]\label{lem:lastplayer}
If a dominating player has an option from which the other player cannot move, then the dominating player will win the game.
\end{lem}
\begin{proof}
Suppose first that the dominating player, say Left, does not hold the marker. Then she goes all in, and moves such that Right cannot move. Since Right still holds the marker, he will lose the game. Namely, at her final bid, Left bids 0, and hence Right wins this final bid. 

Suppose next that the dominating player, say Left, holds the marker. She goes all in, and includes the marker in her bid. She wins the bid and moves such that Right cannot move. Again, the game ends as in the previous paragraph.
\end{proof}

By this lemma, we may use the term {\em last move wins}  with the same meaning as in alternating normal play. Henceforth, we adapt the following wording convention: {\em alternating play} means the classical normal play games \cite{BCG2001}, and {\em bidding play} is our generalization of those games. All our games will be normal play;  we vary the move order convention, but not the winning condition.

\section{A motivating result and a generalization of the impartial game tree}
Standard outcome classes in alternating combinatorial game theory are $$\nL, \nN,\nP,\nR:$$ Left wins, the Next player wins, the Previous player wins, and Right wins, respectively \cite{S2013}. 

Let $G$ be an {\em impartial game}. That is,  for all followers $H$ of $G$, $\HL=\HR$. For impartial games, in alternating play, only the outcomes \nN\ and \nP\ apply. 
The following result about impartial bidding games encouraged us to take a closer look at the full class of partizan bidding games. We will prove a more general result in Theorem~\ref{thm:strongdicot}.
\begin{thm}\label{thm:impartial}
Consider $\tb\in\mathbb N_0$ and an impartial game form $G$. If $2p=\tb$, then Left wins bidding $(G, \hat p)$ if and only if alternating normal play $G$ is an \nN\!-position. If $2p>\tb$, i.e. Left is strictly dominating, then Left wins $(G, \tilde p)$, unless $G=0$, in which case the player who holds the marker loses. 
\end{thm}

\begin{proof}
Omitted.
\end{proof}

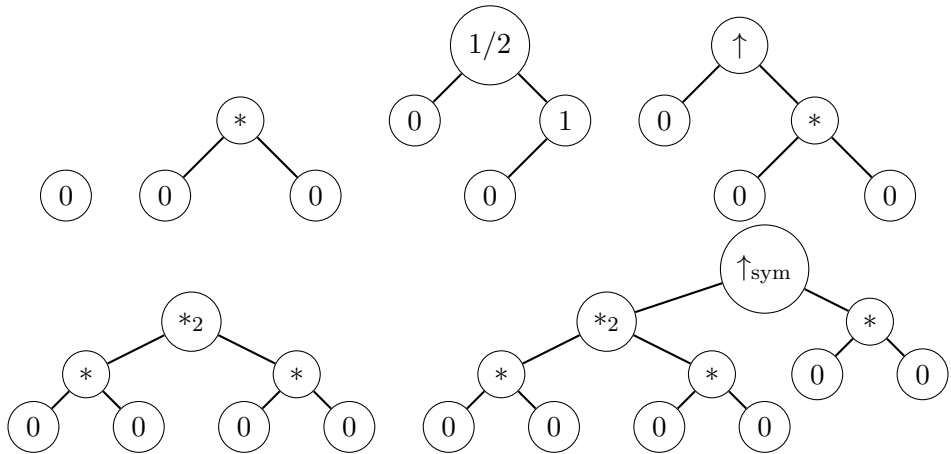
\begin{figure}[ht]
\centering{
\begin{tikzpicture}[scale = 1]
\begin{scope}[every node/.style={circle, draw}]  
    \node (1) at (0,0) {$0$};
    %\node (3) at (1,-1) {$0$};
    %\node (4) at (-1,-1) {$0$};
    %\node (0) at (-1,1) {$*$};
    %\node (1) at (-2,0) {$0$};
    %\node (5) at (-9,0) {5};
    \draw (0, 1) {};%this trick controls the relative vertical distance of tikz pictures
\end{scope}
\begin{scope}[>={Stealth[black]},
              every node/.style={fill=white,circle},
              every edge/.style={draw=black,thick}] 
    %\path [<-] (5) edge[bend left=30] node {3W} (2);
    %\path [-] (1)edge(0);
    %\path [-] (2)edge(0);
    %\path [-] (2) edge (3);
    %\path [-] (2) edge (4);
    %\path [<->] (3) edge[bend left=30] node {3W} (0);
\end{scope}
\end{tikzpicture}\hspace{.5 cm}
\begin{tikzpicture}[scale = 1]
\begin{scope}[every node/.style={circle, draw}]  
    \node (2) at (0,0) {$0$};
    %\node (3) at (1,-1) {$0$};
    %\node (4) at (-1,-1) {$0$};
    \node (0) at (-1,1) {$*$};
    \node (1) at (-2,0) {$0$};
    %\node (5) at (-9,0) {5};
    \draw (0, 1) {};%this trick controls the relative vertical distance of tikz pictures
\end{scope}
\begin{scope}[>={Stealth[black]},
              every node/.style={fill=white,circle},
              every edge/.style={draw=black,thick}] 
    %\path [<-] (5) edge[bend left=30] node {3W} (2);
    \path [-] (1)edge(0);
    \path [-] (2)edge(0);
    %\path [-] (2) edge (3);
    %\path [-] (2) edge (4);
    %\path [<->] (3) edge[bend left=30] node {3W} (0);
\end{scope}
\end{tikzpicture}\hspace{.5 cm}
\begin{tikzpicture}[scale = 1]
\begin{scope}[every node/.style={circle, draw}]  
    \node (2) at (0,0) {$1$};
    %\node (3) at (1,-1) {$0$};
    \node (4) at (-1,-1) {$0$};
    \node (0) at (-1,1) {$1/2$};
    \node (1) at (-2,0) {$0$};
    %\node (5) at (-9,0) {5};
    \draw (0, 1) {};%this trick controls the relative vertical distance of tikz pictures
\end{scope}
\begin{scope}[>={Stealth[black]},
              every node/.style={fill=white,circle},
              every edge/.style={draw=black,thick}] 
    %\path [<-] (5) edge[bend left=30] node {3W} (2);
    \path [-] (1)edge(0);
    \path [-] (2)edge(0);
    %\path [-] (2) edge (3);
    \path [-] (2) edge (4);
    %\path [<->] (3) edge[bend left=30] node {3W} (0);
\end{scope}
\end{tikzpicture}\hspace{.5 cm}
\begin{tikzpicture}[scale = 1]
\begin{scope}[every node/.style={circle, draw}]  
    \node (2) at (0,0) {$*$};
    \node (3) at (1,-1) {$0$};
    \node (4) at (-1,-1) {$0$};
    \node (0) at (-1,1) {$\up$};
    \node (1) at (-2,0) {$0$};
    %\node (5) at (-9,0) {5};
    \draw (0, 1) {};%this trick controls the relative vertical distance of tikz pictures
\end{scope}
\begin{scope}[>={Stealth[black]},
              every node/.style={fill=white,circle},
              every edge/.style={draw=black,thick}] 
    %\path [<-] (5) edge[bend left=30] node {3W} (2);
    \path [-] (1)edge(0);
    \path [-] (2)edge(0);
    \path [-] (2) edge (3);
    \path [-] (2) edge (4);
    %\path [<->] (3) edge[bend left=30] node {3W} (0);
\end{scope}
\end{tikzpicture}
\vspace{.5 cm}
\begin{tikzpicture}[scale = .7]
\begin{scope}[every node/.style={circle, draw}]  
    \node (2) at (2,1) {$*$};
    \node (3) at (3,0) {$0$};
    \node (4) at (-1,0) {$0$};
    \node (0) at (0,2) {$*_2$};
    \node (1) at (-2,1) {$*$};
    \node (5) at (-3,0) {$0$};
     \node (6) at (1,0) {$0$};
    \draw (0, 1) {};%this trick controls the relative vertical distance of tikz pictures
\end{scope}
\begin{scope}[>={Stealth[black]},
              every node/.style={fill=white,circle},
              every edge/.style={draw=black,thick}] 
    %\path [<-] (5) edge[bend left=30] node {3W} (2);
    \path [-] (1)edge(0);
    \path [-] (2)edge(0);
    \path [-] (2) edge (3);
    \path [-] (2) edge (6);
    \path [-] (1) edge (5);
    \path [-] (1) edge (4);
    %\path [<->] (3) edge[bend left=30] node {3W} (0);
\end{scope}
\end{tikzpicture}\hspace{.5 cm}
%%%%%%%%%%%%%%UPSYM%%%%%%%%%%%%%%%%%%%%
\begin{tikzpicture}[scale = .7]
\begin{scope}[every node/.style={circle, draw}]  
    \node (2) at (2,1) {$*$};
    \node (3) at (3,0) {$0$};
    \node (4) at (-1,0) {$0$};
    \node (0) at (0,2) {$*_2$};
    \node (1) at (-2,1) {$*$};
    \node (5) at (-3,0) {$0$};
     \node (6) at (1,0) {$0$};  
    \node (7) at (3,3) {$\up_{\mathrm{sym}}$};
    \node (8) at (5,2) {$*$};
    \node (9) at (4,1) {$0$};
    \node (10) at (6,1) {$0$};
     %\node (6) at (0,-1) {$0$};
    \draw (0, 1) {};%this trick controls the relative vertical distance of tikz pictures
\end{scope}
\begin{scope}[>={Stealth[black]},
              every node/.style={fill=white,circle},
              every edge/.style={draw=black,thick}] 
    %\path [<-] (5) edge[bend left=30] node {3W} (2);
    \path [-] (1)edge(0);
    \path [-] (2)edge(0);
    \path [-] (2) edge (3);
    \path [-] (2) edge (6);
    \path [-] (1) edge (5);
    \path [-] (1) edge (4);   
    \path [-] (0) edge (7);
    \path [-] (7) edge (8);
    \path [-] (8) edge (9);
    \path [-] (8) edge (10);
    %\path [<->] (3) edge[bend left=30] node {3W} (0);
\end{scope}
\end{tikzpicture}\vspace{.2 cm}

}\caption{The literal form games $0$, $*$, $1/2$, $\up$, $*_2=\cg{*}{*}$ and $\up_{\mathrm{sym}}=\cg{*_2}{*}$.
}\label{fig:1}
\end{figure}

\begin{example}
 Figure~\ref{fig:1} displays six game trees, of which the naming of the first four should be familiar to the reader. The two game trees on the bottom row, $*_2$ and $\up_{\mathrm{sym}}$, have non-standard names due to their special interest to this study (note $*_2\ne *2=\cg{0,*}{0,*}$). Let us review some behavior of each of these six games with respect to differences and similarities between alternating play and bidding play. In $0$, which is a $\nP$-position in alternating play, the player without the marker wins. In $*$, which is an $\nN$-position in alternating play, the dominating player wins. It is easy to verify that $1/2$ is a Left win in both alternating play and bidding play (independently of who holds the marker). We know that the game $\up$ is a Left win in alternating play; however, in bidding play, Right can win $\up$, if he holds a sufficient budget (even with $\tb=1$). Observe that, in the game $*_2$ a player needs to win 2 consecutive bids in order to  win the game, which is the index of $*_2$. This can only happen if $\tb>0$; in alternating play, of course $*_2 = 0$,  and does not deserve a special name. The game $\up_{\mathrm{sym}}$ is discussed below using the definition of a {\em symmetric ending game}.
\end{example}
%A player who wins the bid strictly will include the marker if and only if the move results in a game where they have no move.\ur{This requires a two sentence proof.} 

%\section{A generalization of the impartial game tree}

Consider the following generalization of impartial games, to {\em symmetric ending} games. If one of the players has a terminal move, then so has the other player. And if one of the players cannot move then neither can the other player.  That is, $G$ is a symmetric ending game if for all followers $H$, $0\in \HL$ if and  only  if $0\in \HR$, and $\HL=\nil$ if and only if $\HR=\nil$. 

The family of symmetric ending games is larger than the impartial games in that it admits the alternating play outcomes $\mathscr{ L}$ and $\mathscr{R}$. For example, the literal form $\up_{\mathrm{sym}}=\cg{\cg{*}{*}}{*}$ has outcome $\mathscr{L}$. Note that the alternating play equivalent form $\cg{0}{*}=\;\up$ does not have a symmetric ending. %The latter game is an all-small or dicot game, so symmetric ending is smaller than dicot in terms of bidding games.

Recall that the dicot game forms are defined recursively by: if one of the players has a move, then so does the other, and each option is a dicot. The family of dicots contributes a major step from impartial games towards the full family of partizan games. In alternating play, the symmetric ending games are equivalent to the dicots. To see this, note that each Left move to $0$, where Right does not have a move to $0$ (and vice versa) may be replaced with a move to the game $\{ * |*\}$ (this does not hold any longer in generic bidding play). %An example of this was given in the previous paragraph. 
We have the following inclusion diagram:
\begin{center}
    impartial $\subset$ symmetric ending $\subset$ dicot $\subset$ partizan.
\end{center}

The point here is that the symmetric ending condition permits a similar induction proof as that for impartial bidding games (Theorem~\ref{thm:impartial}). 

\begin{thm}\label{thm:strongdicot}
Consider a symmetric ending game $G$, and let $\tb \in\mathbb N_0$.  If $G$ is terminal, then the marker holder loses.  Otherwise the dominating player wins, unless $\tb$ is even and the players have an equal share of the budget. In this case Left wins $(G, \rwhat {\tb/2})$ if and only if alternating play $G$ is an \nl or \nN-position, and Right wins $(G, \tb/2)$ if and only if alternating play $G$ is an \nr or \nN-position. 
\end{thm}

\begin{proof}
In each case, we must present a strategy for the proposed winner. If $G$ is terminal, the non-marker holder  bids $0$ and loses the bid, but wins the game.

Next, consider a non-terminal $(G,\tilde p)$, with $p>\tb/2$.  The strictly dominating player, here Left, bids $0$, until there is a move to a terminal position. By the symmetric ending condition, if one of the players can terminate the game, so can the other. Left goes all in and wins the last move, independently of whether she holds the marker.

For the remainder of the proof, we consider an even total budget $\tb$ and a non-terminal game $(G,\rwhat{\tb/2})$. Suppose  this non-terminal $G$ has a terminal move. Since $G$ is a symmetric ending game, both Left and Right will have a move to a terminal position. Thus $G$ is an  \nN-position in alternating play. And in bidding play, Left, who holds the marker, bids $\rwhat{\tb/2}$, and wins the last move.  %Suppose first that Left has a terminal move. Then $G$ is an \nN-position in alternating normal play, since the symmetric ending condition ensures that if the position is \nl (or \nR), then it does not have a move to a terminal position. Therefore, in the bidding variation, the marker holder, here Left, bids $\hat p$ and wins the last move. 

Suppose next that $G$ is a non-terminal \nr or \nP\!-position in alternating  play;  note that $G$ does not have a terminal move.  We must show that Left will lose the bidding variation. Right, who does not have the marker, bids $0$. If Left bids $0$, then Left wins the move and, by definition of \nr and \nP\!-positions, has to move to an \nr or \nN\!-position, and Right gets the marker. That is losing for Left, by induction. If Left bids $\tilde{\ell} > 0$, Right becomes the dominating player, and wins (by the second paragraph in the proof), because $G^L$ is non-terminal. %Indeed, if $G^L$ were terminal, then alternating play $G\not\in \nR\cup \nP$. 

Suppose next that $G$ is an \nl or \nN-position in alternating play. If $G$ has a terminal move then Left wins by the second paragraph of the proof. Now let us assume $G$ does not have a terminal move.  %, for which there is no move to a terminal position. 
We must prove that Left, who holds the marker, wins. Left bids $0$. If Right also bids 0, then he gets the marker, and Left can move to an alternating play \nl or \nP-position, and win by induction. If Right bids $r>0$, then he will win the bid but Left becomes a dominating player, and so wins by using the second paragraph of the proof, since by assumption no option of $G$ is terminal.

In the game $(G, \tb/2)$ Right holds the marker. Hence by symmetry he wins $(G, \tb/2)$ if and only if alternating play $G$ is an \nr  or \nN-position.
\end{proof}

\section{Outcomes in bidding games}\label{sec:out}
The alternating play outcomes generalize in a word notation, where $\nL=\L\L, \nN=\L\R, \nP=\R\L, \nR=\R\R$. For example ``$\L\R$" means that Left wins when Left starts and Right wins when Right starts. 
This can also be visualized as an outcome (as defined below) for bidding play when $\tb=0$, as discussed in Theorem~\ref{thm:tb0}.  In general, given a game $(\tb, G, \tilde{p})$, the winner is 
%the winner in pure subgame perfect equilibrium, given a current player (the winner of the current bid), is 
either L (Left) or R (Right), with the usual total order $\L>\R$. This induces a partial order of outcomes that generalizes alternating play; see Definition~\ref{def:outrel}. 

For example, with $\tb=1$, one might envision 16 partially ordered candidate outcome classes, in word notation, with the largest (smallest) outcome LLLL (RRRR), and so on. Here the leftmost letter symbolizes the outcome in perfect play when Left has budget 1 and the marker; the second letter corresponds to the result in perfect play when Left has the marker but no budget, and symmetrically for the last two letters. An example of an incomparable pair of outcomes would be LRLR and RLRL.

%That there is indeed exactly one equilibrium needs a proof, since we study pure strategies. A similar result is proved in Bidding cumulative Games [insert citation].

%Now, a partial order of games may be defined in the standard way for combinatorial games, and a major problem is, given any total budget, whether the equivalence class of the neutral element, $0=\cg{\emptyset}{\emptyset}$, is infinite. Recall that in normal play it is rich but in (full) mis\`ere play the unique $0$ is the empty game. We will find an infinite class of $0$-games in Theorem~\ref{thm:infinite}.

%In this case, a study of the strong dicots, may yield a richer structure of equivalence classes of games. 
 %Let us list the outcomes of all games of birthday 2 for $\tb =1$. 

 Our notion of {\em outcome} is a $2(\tb+1)$ tuple of pure subgame perfect equilibria, representing, given any budget partition,  who is the winner in perfect play. 
The pure subgame perfect equilibrium of a game $(\tb, G,\tilde p)$ is denoted by $o(G,\tilde p)=o_\tb(G,\tilde p)$. Sometimes we refer to this as a `partial outcome'.
\begin{defi}[Outcome]
The outcome, $o(G)=o_\tb(G)$, of the game $(\tb, G)$ is $$o(G)=(o(G,\hat\tb),\ldots , o(G,\hat 0),o(G, \tb),\ldots , o(G,0)).$$
Here the first half of the outcome corresponds to when Left holds the marker and the rest corresponds to when Right holds the marker. The length of the outcome is $2(\tb+1)$.
\end{defi}

Since this notation can be quite lengthy, we instead adopt word notation. For example, instead of $(\R,\R,\L,\L)$ we simply write RRLL. 
\begin{defi}[Outcome Relation]\label{def:outrel} Consider a fixed $\tb$ and the set of all budgets $\B$. Then for any games $G$ and $H$, $o(G) \ge o(H)$ if,$~\forall~\tilde p \in \B, o(G, \tilde p) \ge o(H, \tilde p)$.\footnote{It is easy to check that the outcome relation is reflexive, antisymmetric and transitive. Hence the set of all outcomes together with this relation is a poset.}\end{defi}

In alternating play, each zugzwang (a game where no player has an incentive to move) belongs to the outcome class \nP. Here, we define a {\em zugzwang} as: no player wants to win the bid. Therefore, it is not the monetary strength, but the marker ownership, that defines the zugzwang concept; obviously both players will bid $0$, whenever they do not want to move, and the marker will decide who is the current player (i.e. in perfect play, the losing player). Therefore, with $\tb=1$, the only zugzwang is the outcome RRLL, because this corresponds to ``Right wins when Left has the marker'' and ``Left wins when Right has the marker''.  An outcome such as RLRL would be rare, if it appears at all, since Right wins without either money or marker, but loses if he is given a dollar. Next, we prove that such outcomes are impossible; outcomes are {\em monotone}.
\begin{thm}[Outcome Monotonicity]\label{thm:monotone} Consider a fixed $\tilde p\in \B$, with $p<\tb$. Then, for all games $G$, $o(G,\tilde p)\le o(G,\tilde{p+1})$.
\end{thm}
\begin{proof}
If $o(G,\tilde p)=\R$ there is nothing to prove, so suppose $o(G,\tilde p)=\L$. We must prove that $o(G,\tilde{p+1})=\L$. First assume that $\tilde p=p$, i.e. Right holds the marker.\\  

\noindent {\bf Case 1:} Suppose that Left wins $(G,p)$ by bidding $0\le \ell\le p$. Then, she knows that:
\begin{enumerate}[A)]
\item  If $\ell>0$, there is a Left option, $G^L$, such that $o(G^L,p-\ell)=\L$;
\item  For all Right options, $G^R$, $o(G^R,p+r)=\L$, if $r>\ell$;
\item For all Right options, $G^R$, $o(G^R,\rwhat {p+r})=\L$, if $r\ge \ell$. This case is a tie, or Right includes the marker in the bid.
\end{enumerate}
We must present a Left strategy that beats every Right strategy in the game  $(G,  p+1)$. Left bids the same $\ell$ in the game $(G,  p+1)$ as she was bidding in the game $(G, p)$. Then, by induction on the birthday of the game tree:\\ 

\noindent A1) If $\ell > 0$, $o(G^L,p+1-\ell)=\L$, where $G^L$ is the same as in A above;\\ 
\noindent B1) For all $r>\ell$, for all $G^R$, $o(G^R,p+1+r)=\L$;\\ 
\noindent C1) For all $r\ge \ell$, for all $G^R$, $o(G^R,\rwhat {p+1+r})=\L$ (in case of a tie or Right included the marker in the bid). \\

\noindent Note that in case there is no $G^R$, then B, B1, C and C1 are vacuously true.  This proves that $o(G,p+1)=\L$, in case $o(G,p)=\L$.\\

\noindent {\bf Case 2:}  Suppose that Left wins $(G,\hat p)$ by bidding $\tilde \ell \in \B$, $0\le \ell\le p$. Then:
\begin{enumerate}[A)]
\item If $\ell\le q=\tb-p$, there is a Left option, $G^L$, such that $o(G^L,p-\ell)=\L$. That is, Right can tie this bid, in which case Left loses the marker;
\item If $\ell>0$, there is a Left option, $G^L$,  such that $o(G^L,\rwhat{p-\ell}\, )=\L$ or such that $o(G^L,p-\ell)=\L$. Left can choose to include the marker or not to include the marker;
\item If $\ell <q$, then for all Right options, $G^R$, and all $\ell< r\le q$, $o(G^R,\rwhat{p+r})=\L$, if $r>\ell$.
\end{enumerate}

In the game $(G, \rwhat{p+1})$, Left bids the same $\tilde \ell$, as she was bidding in the game $(G, \hat{p})$ and plays the same in case of winning a bid. Right, now has the smaller budget $q-1$. The results follow by induction on the birthday of the game tree; in details:\\ 

\noindent A1) If $\ell\le q-1$, then  if Right ties the bid, Left may play $G^L$ as in A above to get $o(G^L,p+1-\ell)=\L$;\\ 
\noindent B1) If $\ell \ge q$, then Left wins the bid, and with $G^L$ is as in B above, $o(G^L,\rwhat{p+1-\ell})=\L$ or $o(G^L,p+1-\ell)=\L$, where the latter case includes the marker in the bid;\\
\noindent C1) If $\ell<q-1$, then for all $\ell <r \le q-1$, $o(G^R,\rwhat{p+1+r})=\L$.\\ 

This proves that $o(G,\hat p)\le o(G, \rwhat{p+1})$, which together with Case 1, conclude the proof.
%\noindent C) For all Right options, $o(G^R,\hat {p+r})=\L$, if $r\ge \ell$. This case is a tie, or Right includes the marker in the bid;\\ 
 %Left has all available bids from budget $p$. Hence, at each stage of play, the same moves are available in the game on budget $p+1$. Moreover, a zugzwang threat, in case of holding the marker, has the same defense possibilities for budget $p+1$ as for budget $p$. Namely include the marker to the bid in both cases. 
\end{proof}

%Of course, an equivalent statement is: Consider a Left budget $p<\tb$. If $o(G,p+1)=\R$, then $o(G,p)=\R$, and if $o(G,\widehat {p+1})=\R$, then $o_\R(G,\hat p)=\R$.  Yet another way to write this is simply: for all $G$, for all $0<p<\tb$, and fixed $\tilde p$, $o(G,\tilde p)\le o(G,\tilde{p+1})$. 
If these monotone properties hold, the outcome is {\em monotone}.
%\footnote{A player is never forced to become the current player via monetary advantage, since the 0-bid is always available. 
%If one of the players did not have the 0-bid available, the monotonicity may break. Here we study games where all bids $0,\ldots, p$ are available to Left, and analogously for Right.} 

Observe that in our bidding convention, a player is never forced to become the current player via monetary advantage, since the 0-bid is always available. However, if a player with a nonzero budget is not allowed to bid $0$, then monotonicity may break. To see this consider the total budget $\tb=1$, and Right cannot bid $0$, unless his budget is $0$. Then we will have $o(0, \hat{0}) = \L$, but $o(0, \hat{1}) = \R$. Thus, here we study games where if Left's budget is $p$ then all bids $0, \ldots, p$ are available to her, and analogously for Right.
% \begin{obs}
% There are 9 monotone outcome classes for $\tb =1$. And in general, there are $(\tb+2)^2$ monotone outcomes for $\tb\in\mathbb N_0$. 
% \end{obs}

Next, to find the number of monotone outcomes, fix a $\tb\in\mathbb N_0$.
%Then an outcome consists of $2(\tb + 1) $ partial outcomes, in which $\tb + 1$ partial outcomes correspond to when Left holds the marker, which is   $o(G, \hat{\tb}) \cdots o(G, \hat{0})$,  and the rest of them correspond to when Right holds the marker.
Consider the case when Left holds the marker. The only possible monotone sequences of $\tb + 1$ partial outcomes are $\L\L\!\cdots\!\L$, $\L\L\! \cdots\! \L\R$, $\ldots$, $\R\R\!\cdots\!\R$. The number of such monotone sequences of partial outcomes is $\tb+ 2$.

Similarly, when Right holds the marker, we will have  $\tb+ 2$ possible monotone sequences of  $\tb+ 1$ partial outcomes. Thus, from the definition of an outcome, by the independency of such half outcomes there are $(\tb+2)^2$ possible monotone outcomes.

For example, with $\tb=1$, the monotone outcomes are LLLL, LLLR, LLRR, LRLL, LRLR, LRRR, RRLL, RRLR and RRRR. 

\begin{obs}
For $\tb\in\mathbb N_0$ we have $(\tb+2)^2$ monotone outcomes. 
\end{obs}
% \begin{proof}
%     Given a \tb, an outcome consists of $2(\tb + 1) $ partial outcomes, in which $\tb + 1$ partial outcomes correspond to when Left holds the marker and the rest of them correspond to when Right holds marker. 
    
%     Consider the case when Left holds the marker. The possible monotone sequences of $\tb + 1$ partial outcomes are $\L\L\cdots\L$, $\L\L \cdots \L\R$, $\L\L \cdots \L\R\R$, $\ldots$, $\L\R\cdots\R$, $\R\R\cdots\R$. The total number is $\tb+ 2$.
    
%     Similarly we will have  $\tb+ 2$ number of possible monotone sequence of  $\tb+ 1$ partial outcomes when Right has the marker. Thus, in total there are $(\tb+2)^2$ possible monotone outcomes.
% \end{proof}

% For example, with $\tb=1$, the monotone outcomes are LLLL, LLLR, LLRR, LRLL, LRLR, LRRR, RRLL, RRLR and RRRR. 

\begin{table}[ht]
\centering{\caption{Some game forms born by day 2 and their outcomes, for all total budgets $\le 3$.}\label{tab:1}
\begin{tabular}{|c|c|c|c|c|}
\hline
 $G$ & $o_0(G)$ & $o_1(G)$ & $o_2(G)$ & $o_3(G)$   \\ \hline
 $0=\cg{\emptyset}{\emptyset}$ & $\R\,\L$ & $\R\R\,\L\L$ & $\R\R\R\,\L\L\L$ & $\R\R\R\R\,\L\L\L\L$\\ \hline
 $*=\cg{0}{0}$ & L\,R & LR\,LR & LLR\,LRR & LLRR\,LLRR\\ \hline
  $1=\cg{0}{\nil}$ & L\,L & LL\,LL & LLL\,LLL & LLLL\,LLLL\\ \hline
  $\up\,=\cg{0}{*}$ & L\,L & LL\,LR & LLL\,LLR & LLLR\,LLLR\\ \hline
    $\cg{*}{*}$ & R\,L & LR\,LR & LRR\,LLR & LLRR\,LLRR\\ \hline
     $\cg{*}{\nil}$ & R\,L & LR\,LL & LRR\,LLL & LLRR\,LLLL\\ \hline
  $1/2=\cg{0}{1}$ & L\,L & LL\,LL & LLL\,LLL & LLLL\,LLLL\\ \hline
  $\pm 1=\cg{1}{-1}$ & L\,R & LR\,LR & LLR\,LRR & LLRR\,LLRR\\ \hline
\end{tabular}}
\end{table}
Now the question is, do all monotone outcomes appear? Can we find a game form for each $\tb$ and each monotone outcome? For $\tb=0$ all outcomes are trivially monotone, and indeed, all appear by day 1, namely $o(0)=\nP$, $o(1)=\nL$, $o(-1)=\nR$ and $o(*)=\nN$ (see Table~\ref{tab:1}). For $\tb =1$ all monotone outcomes, except LLRR, appear for games born by day 2, which can be seen in Table~\ref{tab:1} by also using symmetry. Observe that for this outcome, in particular a player loses with a dollar budget, but wins with the marker alone. The immediate question is:   
%Let us look at this outcome where a player loses with a dollar budget, but wins with the marker alone. 
Can the marker be worth more than a dollar? The answer is no.

% \begin{thm}
% Consider $\tb=1$. If $o(\hat 0, G) = \L$, then $o(1,G) = \L$.
% \end{thm}
% \begin{proof}
% Suppose that Left wins with marker alone. That is, Left has a good move, if Left wins the bid by Right bidding 0, and Left has a good response to each Right move if Right bids 1. In the first case Right gets both dollar and marker, and in the second case Left gets both dollar and marker. If the first case appears, then Left can win the second game's bid by bidding the dollar, make the same move and the situation is the same as before: Right gets both dollar and marker. Hence Left wins. If the second case appears, where Left has a good response to each Right option, given both dollar and marker, then Left can bid 0, to lose the bid and she will obtain the marker. The situation after Right's best move will be identical, and hence she will not lose. 
% \end{proof}

% Let us generalize this result.

\begin{thm}[Marker Worth]\label{thm:markerworth}
Consider $\tb\in \mathbb N_0$. Then, for all games $G$, $o(G,\hat p) \le o(G,p+1)$.
\end{thm}

\begin{proof}
Assume that Left wins $(G,\hat p)$ by bidding $\tilde \ell$ (otherwise we are done). We claim that Left wins $(G,p+1)$ by bidding $\ell+1$, unless $\ell>\tb-p$, in which case she bids $\ell$. To prove this, we observe the cases for Right's bidding amount. 
\begin{itemize}
    \item [A)] $r> \ell+1$
    \item  [B)] $r=\ell+1$ 
    \item [C)] $r<\ell+1$.
\end{itemize}
 In case of A, Right might bid $r$ or $\hat r$. The case of Right bidding  $\hat r$ will be treated by using instead B. 

In case A, Right bids $r$ and plays to $(G^R,p+1+r)$, if there exists a Right option $G^R\in\GR$ (otherwise we are done). But, by assumption, Left wins $(G^R, \hat{p+r})$, if $r>\ell$, so by induction she wins $(G^R,p+1+r)$. 

In case B, Right wins the bid and plays to $(G^R, \hat{p+1+r})$, if $G^R$ exists (otherwise done). By monotonicity, Left wins this game. Namely, since $r\ge \ell+1$, then $r>\ell$, and by  assumption, Left wins $(G^R, \hat{p+r})$. 

In case C, Left wins the bid and plays to some $(G^L, p-\ell)$. There is a  Left option $G^L$, since by assumption Left wins playing first in $(G, \hat p)$ by bidding $\ell$; even in the case of $\ell=0$ she might win the bid (if Right bids 0), so she has a defence, by playing some $G^L$. Indeed, if Right can tie $r=\ell$, then by assumption, she wins $(G,\hat p)$ by moving to some $(G^L, p-\ell)$. If Right cannot tie, because $\ell>\tb - p$, then bidding $\ell$ or $\hat \ell$ wins the game $(G,\hat p)$ by assumption. In the first case, this results in some game $(G^L,\hat {p-\ell})$, which Left wins.
Left bids $\ell$ in the game $(G,p+1)$ and moves to $(G^L,p-\ell +1)$, which she wins by induction (here we used that $\ell>\tb-p$ implies $\ell >\tb-(p+1)$). In the second case, Left wins the game $(G^L, p-\ell)$ by assumption, and so by monotonicity, she wins $(G^L,p-\ell +1)$. 
%In this case, she could have won the bid by bidding instead $\ell -1$, a bid which we may assume Right can tie. This results in a game $(G^L,p-\ell+1)$. By induction, Left  wins  this game, since Left wins $(G^L,\hat{p-\ell})$. If Left bidding $\hat \ell$ wins, then the argument is as if Right can tie Left's bid. 

Altogether, Left wins $(G, p+1)$ given that Left wins $(G, \hat p)$. 
\end{proof}
%{\color{blue}
 We can view an outcome as a string of $\L's$ and $\R's$. From Outcome Monotonicity and Marker Worth, Theorems~\ref{thm:monotone} and \ref{thm:markerworth}, we see that not all such strings can appear as an outcome of a game. Thus, let us define the notion of a {\em feasible} outcome. 
%\pr{I feel we should put some writing here.}
\begin{defi}[Feasible Outcome]\label{def:Foutcome}
An outcome is feasible if it satisfies Outcome Monotonicity (Theorem~\ref{thm:monotone}) and Marker Worth (Theorem~\ref{thm:markerworth}). For a given $\tb$, the set of all feasible outcomes is  $\mathcal{O}=\mathcal{O}_\tb$.
\end{defi}

It can be convenient to think of an outcome, without the mention of an underlying game. Thus, the relevance of the set $\mathcal{O}$.
%}

Let us count the number of feasible outcomes for a given $\tb$. Let $a$ be the lowest Left budget for which $o(G, \hat{a}) = \L$ and $a=\tb+1$ if no such $a \in \{0, \ldots, \tb\}$ exists. When Left holds the marker, by Outcome Monotonicity, corresponding to each $a$ there is only one  half-outcome. Next, when Right holds the marker, by Marker Worth, for any $a$, $\max(\tb-a,0)$ number of half-outcomes are not possible. Thus if we denote the triangular numbers by $\Delta (n) = 1+\cdots +n$, then for a given total budget $\tb$, there are $(\tb+2)^2-\Delta(\tb)$ feasible outcomes.

%We denote the triangular numbers by $\Delta (n) = 1+\cdots +n$. 

\begin{obs}
For a given total budget $\tb$, there are $(\tb+2)^2-\Delta(\tb)$ feasible outcomes.
\end{obs}
% \begin{proof}
% Simple counting will do. \pr{comment: needs to write the proof}
% \end{proof}

A challenge will be to demonstrate that all feasible outcomes appear for some game form. We resolve this in Theorem~\ref{thm:main}.

\section{Do all feasible outcomes appear?}\label{sec:outappear}
Now we see that, for $\tb=1$, all feasible outcomes appear by day 2, namely take the rows 1, 2, 3, 4 and 6 in Table~\ref{tab:1} and include the conjugates\footnote{An outcome-conjugate is obtained when the Ls and Rs are reversed and swapped symmetrically.}  of rows 3, 4 and 6. What about the outcomes for $\tb =2$? There are 13 feasible outcome classes. But, will games born by day 2 suffice? For example, what about the feasible outcome LLRLLL? The `nearest' in Table~\ref{tab:1} is $o_2(\cg{*}{\nil})=$ LRRLLL. It turns out that a solution is a game of day 3; namely $\cg{\up\;}{\nil}$ has the desired outcome $o_2(\cg{\up\;}{\nil})=\L\L\R\L\L\L$.  Here Right wins if and only if he has budget $\$2$ and no marker. Clearly, if Right has the marker he loses. If he does not have the marker, he will pass, and so Left will play to $\up\; =\cg{0}{*}$ and hand over the marker. Now, he wins if and only if he wins two consecutive bids, which is possible if and only if he has budget $\hat 2$.

Another feasible outcome that does not appear in Table~\ref{tab:1} is LRRLRR. Again, a game of day 3 has this outcome, 
namely $o_2(\cg{*}{\;\down})=\L\R\R\L\R\R$. That is,
Left wins if and only if she has budget $p=2$, with or without the marker. With budget $2$, she starts by bidding $0$, and wins the marker. Right has to move to down. Now Left has $\hat 2$, and wins the next two bids to win the game. If Left starts with budget $\hat 2$, then she wins two bids, moving to $*$ and then $0$. Right wins any other game. When Right has budget $1$, he begins by bidding 0\, and gets either the marker or another dollar. Thus he wins after Left's move to $*$. If he starts with $\hat 1$, then he begins by bidding 0. If Left bids 0, he plays to $(\down\, , \hat 1\,)$. Now Right will again bid 0 and take the game to either $(*,1)$ or $(*, 0)$ or $(*, \hat{0})$, where he is the dominating player and wins using Theorem~\ref{thm:strongdicot}. Thus Left cannot afford to lose the first move, and so she bids $1$. Right gets all budget and wins from $(*,0)$. The other cases follow by monotonicity.

Together with the games in Table~\ref{tab:1} and conjugate forms, we have now found games for all outcomes with $\tb=2$. Below, in Table~\ref{tab:2}, we will find game forms for all budget partitions with $\tb = 3$. 
Greater total budgets offer more challenges; however, in Theorem~\ref{thm:main} we will provide a general construction. But before that we will define two key tools of our game construction: {\em short form feasible outcome} (Definition~\ref{def:shortform}) and {\em terminating Left budget pair} (Definition~\ref{def:TLBP}). 
\begin{table}[ht!]
\centering{\caption{Game forms for all feasible outcomes, up to conjugation,  with $\tb=3$. See Example~\ref{ex:tab2}. }\label{tab:2}
\begin{tabular}{|l|c|c|}
\hline
Feasible outcome & Shortform & Game \\ \hline
$\L\L\L\L\,\L\L\L\L$& $(0,0)$ & 1\\\hline
$\L\L\L\L\,\L\L\L\R$&$(0,1)$&$\cg{0}{\,\up}$\\\hline
$\L\L\L\R\,\L\L\L\L$&$(1,0)$&$\cg{\cg{0}{\;\up}}{\nil}$\\\hline
$\L\L\L\R\,\L\L\L\R$&$(1,1)$&$\up$\\\hline
$\L\L\L\R\,\L\L\R\R$&$(1,2)$&$\cg{0}{\cg{\down\;}{0}}$\\\hline
$\L\L\R\R\,\L\L\L\L$&$(2,0)$&$\cg{*}{\nil}$\\\hline
$\L\L\R\R\,\L\L\L\R$&$(2,1)$&$\cg{*}{\;\up}$\\\hline
$\L\L\R\R\,\L\L\R\R$&$(2,2)$&$*$\\\hline
%$\L\L\R\R\,\L\R\R\R$&$(2,3)$&\\\hline
$\L\R\R\R\,\L\L\L\L$&$(3,0)$&$\cg{\down\;}{\nil}$\\\hline
$\L\R\R\R\,\L\L\L\R$&$(3,1)$&$\cg{\down\;}{\;\up}$\\\hline
$\R\R\R\R\,\L\L\L\L$& $(4,0)$ & 0\\\hline
\end{tabular}}
\end{table}
\begin{defi}[Short Form Feasible Outcome]\label{def:shortform}
    The short form of a feasible outcome is $(a,b)\in \mathcal{O}$.\footnote{Technically, the short form is a function from words to pairs of nonnegative integers. In practice, henceforth we will only use the short form of a feasible outcome.} Here $0\le a\le \tb+1$ and $0\le b\le \tb+1$ is the smallest Left budget for which she wins, when Left has the marker and does not have the marker respectively. 
\end{defi}

 Consider a short form $(a,b)\in \mathcal{O}$. Then, by Theorem~\ref{thm:markerworth}, 
\begin{align}\label{eq:ba1}
    b\le a+1. 
\end{align}

Recall, an outcome-conjugate is obtained when the Ls and Rs are reversed and swapped symmetrically. Hence, a conjugate short form of $(a,b)$ is $(\tb+1-b, \tb+1-a)$. Note that outcome-conjugation preserves feasibility. 

\begin{example}\label{ex:tab2}
In Table~\ref{tab:2}, the conjugate short forms of the outcomes $$(0, 0), (0,1), (1,0), (1,1), (1,2), (2,0), (2,1), (3,0)$$ are $$(4,4), (3,4), (4,3), (3,3), (2,3), (4,2), (3,2), (4,1)$$ respectively. 
Further, the three outcomes $(2,2), (3,1), (4,0)$ are their own conjugates. 
\end{example}

%\subsection{The game construction}
%\pr{Let us introduce the notion of {\em termination of Left budget pair}. We will use this notion in our construction to enforce the desired outcome.}
We support the next definition with an example. 

\begin{example}
 Consider a sufficient combinatorial game $G$ with $\tb = 50$. Let us explore two distinct scenarios: one where Left begins with a budget of $1$ along with the marker, and the other where she starts with a budget of $2$ also with the marker. Let us say $p_0 = 1$ and $p_0'=2$. Assume that at each bidding stage, the dominating player must outbid the other player. Since Right is the dominant player in both scenarios, he consistently outbids Left in both situations until she becomes dominant in at least one of the scenarios. We get the sequence of Left budget pairs as $(p_1, p_1') = (3,5)$, $(p_2, p_2') = (7,11)$, $(p_3, p_3') = (15,23)$, $(p_4, p_4') = (31,47)$.  Observe that for the $5^{\mathrm{th}}$ round of bidding, Left is the dominating player with the marker in both sequences. Now, Left will start outbidding Right. We get $(p_5, p_5') = (12,44)$ with Right holding the marker in both sequences. Here, observe that Right became the dominating player in the sequence starting from $p_0 = 1$; however, Left remains the dominating player in the sequence starting from $p_0' = 2$.  The dominating player shifts in exactly one of the
sequences at this $5^{\mathrm{th}}$ round of bidding. We will terminate this sequence here, and we will say that this {\em sequence of Left budget pairs terminates} at the   $5^{\mathrm{th}}$ round of bidding. %$\tau=5$.% of bidding and hence $\tau=5$.
\end{example}
%Let us now formally define the concept of a "terminating Left budget pair" and proceed to establish its termination within a finite number of bids.
For any given $\tb$ and any feasible outcome, using the process in this example, we will construct a game form with this outcome. 

\begin{defi}[Terminating Left Budget Pair]\label{def:TLBP}
   Consider a total budget $\tb$ and a sufficiently large combinatorial game form $G$. Consider two Left budgets $p_0$ and $p_0'$ such that $p_0<p'_0$. For both games $(G,p_0)$ and $(G,p_0')$, fix the marker holder (either Left or Right) and assume that one of the players is dominating in both. Assume that in both these games, at each bidding stage, the dominating player outbids the other player. If the dominating player holds the marker, they bid the opponent's budget together with the marker, and otherwise, they bid the minimum amount to outbid the opponent, i.e. the opponent's budget $+1$.  For the games $(G,p_0)$ and $(G,p_0')$, this bidding process generates {\em base sequences} of Left budgets $(p_i)_{i\in I}$ and $(p_i')_{i\in I}$, respectively, where $I$ is a set of indices starting from $0$. By combining these two sequences, we obtain a {\em sequence of pairs of Left budgets} $((p_i,p'_i))_{i\in I}$.  This sequence {\em terminates}   if there is a smallest index, say $\tau$, such that at the $\tau^{\mathrm{th}}$ round of bidding, the dominating player shifts in exactly one of the base sequences.
\end{defi}

In Lemma~\ref{lem:terminate}, we prove that $\tau$ is finite for any given total budget. Additionally, we show that, at termination, Left becomes the dominating player in the base sequence that started with her larger budget.
% \begin{example}
%  Consider an appropriate combinatorial game $G$ with $\tb = 50$. Consider a pair of bidding sequences that starts at the Left budgets $2$ and $1$, respectively with Left holding the marker in both the sequences. Here we have $p_0 = 1$ and $p_0' = 2$.
 
%  Now Right is strictly dominating in both the sequences, thus Right will outbid Left until she becomes dominating in at least one of the sequences. We get $(p_1, p_1') = (3,5)$, $(p_2, p_2') = (7,11)$, $(p_3, p_3') = (15,23)$, $(p_4, p_4') = (31,47)$. Observe that after the $4^{\mathrm{th}}$ round of bidding, Left is the dominating player in both sequences with corresponding Left budgets $(p_4, p_4') = (31,47)$, and where  Left holds the marker in both sequences. Thus, in the $1^{\mathrm{st}}$ shift, the sequence did not terminate.
 
%  Now, Left will start outbidding Right. Then we will have $(p_5, p_5') = (12,44)$ with Right having marker in both sequences. Here observe that Right became the dominating player in the sequence starting from $p_0 = 1$; however Left is the dominating player in the sequence starting from $p_0' = 2$. Thus this sequence of Left budget pairs terminates at $\tau=5$ bidding rounds.% of bidding and hence $\tau=5$.
% \end{example}

\begin{lem}\label{lem:terminate}
For any sufficiently large combinatorial game form, consider a total budget $\tb$ and a pair of Left budgets, $p_0<p'_0$, say. Fix the marker holder. 
\begin{enumerate}[(i)]
    \item There exists a smallest index $\tau$ such that at the $\tau^{th}$ round of bidding, the sequence of pairs of Left budgets $((p_i,p'_i))_{i\in I}$ terminates. 
    %\done{And $\tau$ is the smallest index such that} 
    \item At termination, Left is the dominating player with $p'_{\tau}$ and Right is the dominating player with  $\tb - p_{\tau}$.
\end{enumerate}
%The index set $I=\{0,\ldots ,\tau\}$ is finite, 
\end{lem}

\begin{proof} To prove (i), suppose first that Left holds the marker and Right is the dominating player in both sequences. The first sequence,  which starts with Left budget $p_0$,  will flip dominating player at some point, say after $i>0$ bids. Then the corresponding Left budget will be $p_i= 2p_{i-1}+1 = 2^ip_0+2^{i}-1$. If the second sequence shifts its dominating player before this stage, we are done. 
Also note that in the second sequence the dominating player must already have shifted at this stage. Hence suppose that the second sequence flipped dominating player at the same stage $i$, with Left having the budget $p_i'=2^ip_0'+2^{i}-1$. %The difference between this pair of Left budgets is $2^i(p'-p)\ge 2^i$. 
It will be sufficient to compare the budgets of non dominating player between the shifts. 
At this point, Left has become the dominating player in both sequences. The non-dominating player, Right, has a budget pair with the difference of $|(\tb-p_i)-(\tb-p_i')|= 2^i|(p_0'-p_0)|> |p_0'-p_0)| ; i\geq 1  $. Thus, the absolute difference of budgets of non-dominating player  is strictly increasing at each stage when the dominating player is shifting. Therefore this will not happen forever, since $\tb$ is fixed.

 In case Right is the dominating player but Left does not hold the marker in both sequences, Right can tie Left's budget at the first stage. If the dominating player shifts in exactly one of the sequences, then we are done. However, if it did not, then the proof follows as in the previous paragraph. The other case is when Left is the dominating player, which is symmetric to the cases when Right is the dominating player. Thus the sequence of Left budget pairs  terminates after a finite number of bidding rounds, say $\tau$. %Hence, the index set $I=\{0,\ldots ,\tau\}$ is finite.

For (ii), we begin by proving that, for all possible pairs of Left budgets $(p_i, p_i')$ such that $p_i<p_i'$, we will have $p_{i+1}< p_{i+1}'$. There are four cases. In both sequences, 
\begin{enumerate}[{Case~}1)]
    \item Right is the dominating player but the marker is with Left.   In this case after the first round of bidding, Left will have the marker in both  sequences and $(p_{i+1}, p_{i+1}') = (2p_i+1, 2p_i'+1)$. Clearly $p_{i+1}< p_{i+1}'$, since  $p_i<p_i'$.
    
    \item Right is the dominating player and  the marker is with Right. In this case after the first round of bidding, Left will have the marker in both sequences and $(p_{i+1}, p_{i+1}') = (2p_i, 2p_i')$. In this case also $p_{i+1}< p_{i+1}'$, since  $p_i<p_i'$.
    
    \item Left is the dominating player and  the marker is with Left. In this case after the first round of bidding, Left will not have the marker in both sequences but $(p_{i+1}, p_{i+1}') = (2p_i-\tb, 2p_i'-\tb)$. In this case also $p_{i+1}< p_{i+1}'$, since  $p_i<p_i'$.
    
    \item Left is the dominating player but the marker is with Right.  In this case after the first round of bidding, Left will not have the marker in both sequences but $(p_{i+1}, p_{i+1}') = (2p_i-\tb-1, 2p_i'-\tb-1)$. In this case also $p_{i+1}< p_{i+1}'$, since  $p_i<p_i'$.
\end{enumerate}
By (i), the sequence of Left budget pairs terminates at the $\tau^{\mathrm{th}}$ round of bidding, and using the four cases, recursively, we get $p_\tau<p_\tau'$. Since termination means that there is a shift of dominating player in exactly one of the base sequences, Left must be dominating with $p_\tau'$, and Right must be dominating with  $\tb - p_\tau$.
\end{proof}

%\subsection{The proof of the main theorem}
From the proof of Lemma~\ref{lem:terminate} it is clear that, given $\tb$, $\tau $ is uniquely defined by the pair $(p_0, p_0')$, and it is independent of the given game form. 

Every game has a feasible outcome, i.e. it satisfies Outcome Monotonicity and Marker Worth. Let us now prove the other direction.
\begin{thm}[Main Theorem]\label{thm:main}
Consider any total budget $\tb\in \mathbb N_0$. For all $\omega \in \mathcal{O}$, the set of all feasible outcomes, there is a game form $G$ such that  $o(G)=\omega$. 
\end{thm}
\begin{proof}

Given any total budget $\tb\in \mathbb N_0$, consider an  $\omega \in \mathcal{O}$. Consider the short form $(a,b)$ of $\omega$ (Definition~\ref{def:shortform}). Recall $o_\tb(G,\hat p)=\L$ if and only if $p\ge a$ and $o_\tb(G,p)=\L$ if and only if $p\ge b$. We will build a game form $G$ such that $o_\tb(G)=(a,b)$.  Thus, by Outcome Monotonicity, it suffices to construct a game form $G$ such that 
\begin{equation}\label{eq:game}
\begin{aligned}
&&o(G, \hat{a-1}) = \R, &~ o(G, \hat{a}) = \L, \text{\ and}\\
&&o(G, b-1) = \R, &~  o(G, b) = \L.
\end{aligned}
\end{equation}

The construction of the game form $G$ will rely on both players beginning with 0 bids (see Claim~1). Then, by Outcome Monotonicity, the construction of $G$ according to $\eqref{eq:game}$ reduces to constructing $G^L $ and $G^R$ with the  properties: 
\begin{equation}\label{eq:reducedgame}
\begin{aligned}
&&o(G^L, a-1) = \R, &~ o(G^L, a) = \L,\text{\ and} \\
&&o(G^R, \hat{b-1})= \R , &~ o(G^R, \hat b) = \L. 
\end{aligned}
\end{equation}

\textbf{Claim 1:} If $G$ satisfies  \eqref{eq:reducedgame}, then the player without the marker in \eqref{eq:game} cannot benefit by instead outbidding the opponent.\\

\noindent{\em Proof of Claim 1:}
Suppose our constructed game $G$ satisfies \eqref{eq:reducedgame}. Consider the case when Left has the marker. From \eqref{eq:game} we should have $o(G, \hat a) = \L$. Now suppose, in the game $(G, \hat a)$, that Right instead outbids Left by bidding 1 dollar.  The game becomes $(G^R, \hat{a+1})$. By \eqref{eq:ba1}, we know that $a+1\geq b$ and from \eqref{eq:reducedgame} we have $o(G^R, \hat b) = \L$. This implies $o(G^R, \hat{a+1}) = \L$. 

Thus, when Left has the marker, Right would not benefit by instead outbidding Left. Similarly when Right has the marker Left will not benefit by instead outbidding Right. This proves the claim.\\ 

Now we can go ahead with the construction of the game forms $G^L$ and $G^R$, satisfying ~\eqref{eq:reducedgame}, instead of $G$ satisfying \eqref{eq:game}. The intuition of the construction is as follows: We may assume that a dominating player outbids the other player whenever they face a threat of an imminent loss. We will design terminal threats, i.e. options to the game $0$, for those instances, where a currently non-dominating player must win the game. Such threats will eventually flip the dominating player, and at this point we wish to let the player who must win, terminate the game. However, it may happen that the dominating player flips simultaneously in both game sequences, say  those initiated by the respective Left starting budgets $p'\ge a$ and $p\le a-1$. In this case, we cannot yet let the currently dominating player, say Left, have a terminating option, but we will rely on Lemma~\ref{lem:terminate}. The game will proceed with Left, who is now dominating, facing terminal threats. She must outbid Right, until at least one of the bidding sequences flips the dominating player. If exactly one sequence flips, then we terminate the game, and otherwise, repeat. Indeed Lemma~\ref{lem:terminate} assures that this process will finish in a finite number of steps. Moreover it also assures that, after termination, Left will be the dominating player in the sequence starting with Left budget $p'\geq a$ and Right  will be the dominating player in the sequence starting with Left budget $p\leq a-1$. See Figure~\ref{fig:G0} for an example of such a constructed game tree.

\begin{figure}[ht]
\centering{
\begin{tikzpicture}[scale = .9]
\begin{scope}[every node/.style={circle, fill=white,inner sep=1, draw}]  
    \node (0) at (0,0) {$G_0$};;
    \node (1) at (1,-1) {$G_1$};
    \node (2) at (2,-2) {$G_2$};
    \node (3) at (3,-3) {$G_3$};
    \node (4) at (4,-4) {0};
    \node (5) at (-1,-1) {0};
    \node (6) at (0,-2) {0};
    \node (7) at (1,-3) {0};
    \node (8) at (2,-4) {$G_4$};
    \node (9) at (1,-5) {$G_5$};
    \node (10) at (0,-6) {0};
    \node (11) at (2,-6) {$*$};
    \node (12) at (3,-5) {0};
    \node (13) at (3,-7) {0};
    \node (14) at (1,-7) {0};
    \draw (0, 1) {};%this trick controls the relative vertical distance of tikz pictures
\end{scope}
\begin{scope}[>={Stealth[black]},
             every edge/.style={draw=black,thick}] 
    \path [-] (1)edge(0);
    \path [-] (2)edge(1);
    \path [-] (3) edge (2);
    \path [-] (4) edge (3);
    \path [-] (5) edge (0);
    \path [-] (6) edge (1);
    \path [-] (7) edge (2);
    \path [-] (3) edge (8);
    \path [-] (11) edge (13);
    \path [-] (11) edge (14);
    \path [-] (8) edge (12);
    \path [-] (10) edge (9);
    \path [-] (8) edge (9);
    \path [-] (11) edge (9);
\end{scope}
\end{tikzpicture}\caption{Given a feasible outcome $(a,b)$ on total budget $\tb$, a Left option $G^L=G_0$ has been constructed such that $o(G)=(a,b)$. We study the partial outcomes on the Left budgets $p=a-1$ and $p'=a$. Right begins by a pass, and then he outbids Left three times. At this point the dominating player shift. Eventually the game will end by one of the players getting the last move in $*$.}\label{fig:G0}
}
\end{figure}
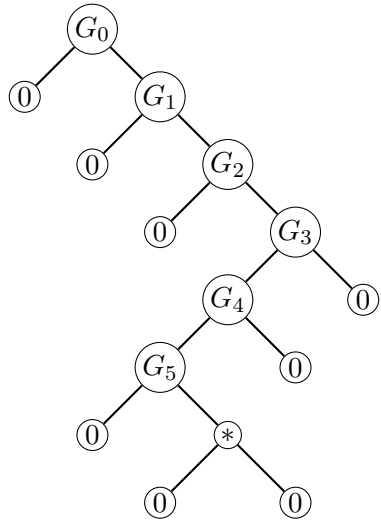
Let us now detail the construction of $G^L = G_0$ (say). We use a pair of sequences $(p_i)$ and $(p'_i)$ of monetary Left-budgets,  starting with $p_0=a-1$ and $p'_0=a$, as required in \eqref{eq:reducedgame}, while we define a sequence of followers, $(G_i)$, of $G$, depending on the threshold $a$ alone. We study different cases of the Left budget $a$.
\begin{enumerate}[{Case~}1:]
        \item Consider  $a=\lfloor\tb/2\rfloor+ 1$. Left starts as the dominating player in precisely one of the games, namely in $(G^L,a)$. We let $G_0=G^L=*$.  Then, using Theorem~\ref{thm:strongdicot}, the dominating player will get the last move and win, and thus $\L=o(G^L,a)>o(G^L,a-1)=\R$.
        \item Consider $a>\lfloor\tb/2\rfloor+1$. Left starts as the dominating player in both game sequences. But we must satisfy $\L=o(G^L,a)>o(G^L,a-1)=\R$. To get this, we can use the setting of Lemma~\ref{lem:terminate}. Let's assume  $p_0=a-1$ and $p_0' = a$. We now create threats to the dominating player by giving the opponent a move to $0$, so that the dominating player is forced to outbid the opponent. Here we start by  creating a threat to Left by giving Right a move to $0$, i.e. set $G_0^R = 0$. Hence Left must strictly outbid Right's budget (Right holds the marker). This creates the necessary condition to apply Lemma~\ref{lem:terminate}. Thus, to use Lemma~\ref{lem:terminate}, we will keep creating these threats until the sequence of pairs of Left budgets gets terminated.  Lemma~\ref{lem:terminate} ensures that this will take only finitely many rounds of bidding, say $\tau$. Let $G_{\tau}$ be the game form reached after $\tau$ such rounds of bidding. The Left budgets starting from $(p_0, p_0')$ will become $(p_{\tau}, p_{\tau}')$ respectively, where Left will be dominating with $p_{\tau}'$ but Right will be dominating with \tb-$p_{\tau}$. We will now let $G_{\tau}=*$ to ensure $\L=o(G_{\tau},p_{\tau}')>o(G_{\tau},p_{\tau})=\R$. Thus, we will get $\L=o(G^L,a)>o(G^L,a-1)=\R$ 
        \item Consider $a<\lfloor\tb/2\rfloor+1$. Right starts as the dominating player in both sequences. But we must satisfy $\L=o(G,a)>o(G,a-1)=\R$. We will use the same strategy as in Case 2, to get the desired result.
    \end{enumerate}
The construction of $G^R$ to satisfy \eqref{eq:reducedgame} when Left holds the marker is symmetric.
This completes the construction of the game form according to $\eqref{eq:reducedgame}$ and hence according to $\eqref{eq:game}$.
\end{proof}
%Our construction provides an upper bound on birthdays of game forms with outcomes in $\mathcal{O}_\tb$.
Given a feasible outcome, what birthday of game form do we require?
%The next question is, given a total budget $\tb$, what birthday of game tree is sufficient to get all game forms?
Consider a given total budget $\tb$ and a short form feasible outcome $(a,b)$. Suppose that at the end of the $\tau_a^\mathrm{th}$ and $\tau_b^\mathrm{th}$ rounds of bidding, the Left budgets terminate corresponding to when Left has the marker and does not have the marker, respectively. Then for the feasible outcome $(a,b)$, our constructed game form will have birthday  $\max(\tau_a+2, \tau_b+2)$. 
Hence, by birthdays of game trees no more than $\max_{(a,b) \in \mathcal{O}_\tb}(\max(\tau_a+2, \tau_b+2))$, we can find game forms for  all outcomes in $\mathcal{O}_\tb$.
%Thus an upper bound on the birthdays of game trees with outcomes in $\mathcal{O}_\tb$ is $\max_{(a,b) \in \mathcal{O}_\tb}(\max(\tau_a+2, \tau_b+2))$.

\section{Outcome lattices}\label{sec:lattice}
At this point, one might wonder: given a $\tb$, what exactly are the structures of the feasible outcomes?  Definition~\ref{def:outrel} of Outcome Relation gives rise to a partial order among the set of all outcomes. Proceeding with this idea, by Definitions~\ref{def:Foutcome} and~\ref{def:outrel}, for any $\tb$ the set of all feasible outcomes together with the outcome relation, $(\mathcal{O}_\tb, \ge)$, forms a poset. Observe that the greatest element in this poset is $\L\cdots \L$ (with short form $(0,0)$) and the least element is $\R\cdots \R$ (with short form $(\tb+1, \tb+1)$). In Theorem~\ref{thm:lattice}, we prove that $(\mathcal{O}_\tb, \ge)$ is a lattice.

\begin{figure}[ht!]
\centering{
\begin{tikzpicture}[scale = 1.8]
\begin{scope}[every node/.style={circle, draw}]  
    \node (2) at (0,0) {$1,0$};
    %\node (3) at (1,-1) {N};
    \node (4) at (-1,-1) {$1,1$};
    \node (0) at (-1,1) {$0,0$};
    \node (1) at (-2,0) {$0,1$};
    \draw (0, 1) {};%this trick controls the relative vertical distance of tikz pictures
\end{scope}
\begin{scope}[>={Stealth[black]},
              every node/.style={fill=white,circle},
              every edge/.style={draw=black,thick}] 
    \path [-] (1)edge(0);
    \path [-] (2)edge(0);
    \path [-] (1) edge (4);
    \path [-] (2) edge (4);
\end{scope}
\end{tikzpicture}\hspace{1 cm}
\begin{tikzpicture}[scale = 1.8]
\begin{scope}[every node/.style={circle, draw}]  
    \node (0) at (0,-2) {2,2};
    \node (1) at (0,2) {0,0};
    \node (2) at (-1,-1) {1,2};
    \node (3) at (-1,0) {1,1};
    \node (4) at (-1,1) {0,1};
    \node (5) at (1,-1) {2,1};
    \node (6) at (1,0) {2,0};
    \node (7) at (1,1) {1,0};
    \draw (0, 1) {};%this trick controls the relative vertical distance of tikz pictures
\end{scope}
\begin{scope}[>={Stealth[black]},
              every node/.style={fill=white,circle},
              every edge/.style={draw=black,thick}] 
    \path [-] (7)edge(3);
    \path [-] (5)edge(3);
    \path [-] (1) edge (7);
    \path [-] (1) edge (4);
     \path [-] (4)edge(3);
    \path [-] (2)edge(3);
    \path [-] (6) edge (7);
    \path [-] (5) edge (6);
    \path [-] (0) edge (2);
    \path [-] (5) edge (0);
\end{scope}
\end{tikzpicture}

}\caption{The left picture displays the alternating play outcome diamond, and the right picture is the Hasse diagram of outcomes for $\tb=1$. The notation is in the short form of a game: the smallest Left budget for which she wins, with and without marker respectively.}\label{fig:lattices}
\end{figure}
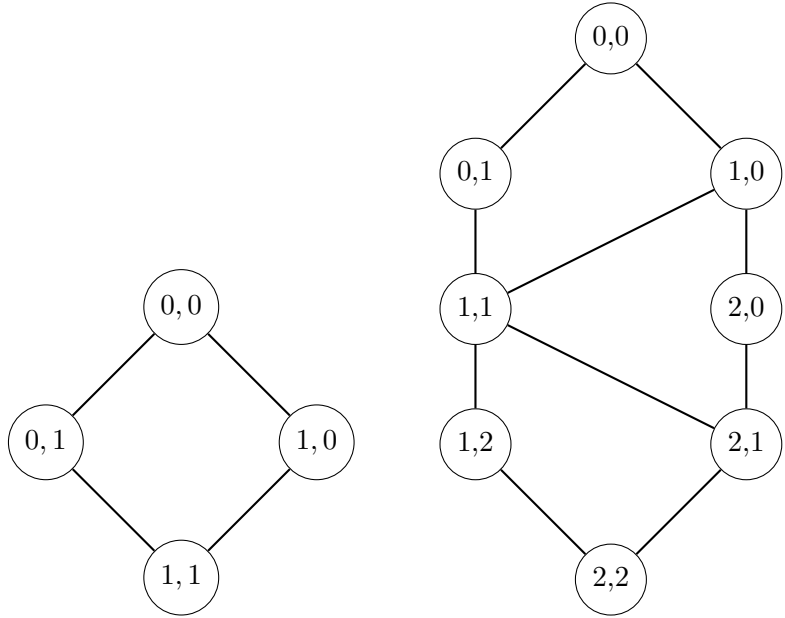

Let us now provide a recursive construction of its Hasse diagram. Recall the outcome diamond for normal play, i.e. $\tb = 0$: $\L\L>\L\R>\R\R$ and  $\L\L>\R\L>\R\R$, with $\L\R \ngeqslant\R\L$ and  $\R \L\ngeqslant\L\R$. This is displayed to the left in Figure~\ref{fig:lattices} by using the short form notation; the Hasse diagram for $\tb=1$ is displayed to the right.  

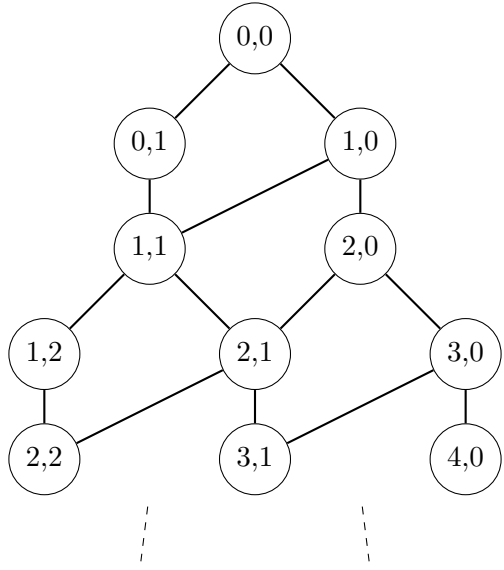
\begin{figure}[ht!]
\centering{
\vspace{.5 cm}
\begin{tikzpicture}[scale = 1.4]
\begin{scope}[every node/.style={circle, draw}]  
    \node (1) at (0,2) {0,0};
    
    \node (4) at (-1,1) {0,1};
    \node (7) at (1,1) {1,0};
    
    \node (3) at (-1,0) {1,1};
    \node (6) at (1,0) {2,0};

    \node (2) at (-2,-1) {1,2};
    \node (5) at (0,-1) {2,1};
    \node (0) at (2,-1) {3,0};
    
     \node (8) at (-2,-2) {2,2};
    \node (9) at (0,-2) {3,1};
    \node (10) at (2,-2) {4,0};
    \draw (0, 1) {};%this trick controls the relative vertical distance of tikz pictures
\end{scope}
\begin{scope}[>={Stealth[black]},
              every node/.style={fill=white,circle},
              every edge/.style={draw=black,thick}] 
    \path [-] (7)edge(3);
    \path [-] (5)edge(3);
    \path [-] (1) edge (7);
    \path [-] (1) edge (4);
     \path [-] (4)edge(3);
    \path [-] (2)edge(3);
    \path [-] (6) edge (7);
    \path [-] (0) edge (6);
    \path [-] (6) edge (5);
    
     \path [-] (2) edge (8);
    \path [-] (5) edge (9);
    \path [-] (0) edge (10);
    \path [-] (5) edge (8);
    \path [-] (0) edge (9);
    
    \draw (1,-2.3) node {} -- (1.1,-3.1) [dashed] node {};
    \draw (-1,-2.3) node {} -- (-1.1,-3.1) [dashed] node {};
\end{scope}
\end{tikzpicture}

}\caption{The Hasse diagram for $\Gamma$.
%infinite join-semilattice $\Gamma$   generates every lattice of feasible outcomes
}\label{fig:semilattice}
\end{figure}

Let us now construct Hasse diagrams for larger total budgets. Consider a set $V$ of pairs of integers $(a,b)$ with $a,b\in \mathbb N_0$ where $b\le a+1$. Consider a digraph $\Gamma$ on the set $V$, with an upward directed edge from $(a',b')$ to $(a,b)$ if $a+b+1=a'+b'$ and where $a\le a'$ and $b\le b'$. This is displayed in Figure~\ref{fig:semilattice}. These directed edges produce the partial order in the set $V$.
Now for any $\tb \in \mathbb N_0$, we construct a digraph $\Gamma_\tb$ by including the nodes $(a, b)$ for which $a+b \le \tb+1$, together with the nodes $(\tb+1-a,\tb+1-b)$ where $a+b\le \tb$. And $\Gamma_\tb$ inherits the directed edges from $\Gamma$. For an example, see $\Gamma_1$ in the picture to the right in Figure~\ref{fig:lattices}. 

\begin{thm}\label{thm:digraphlattice}
    For all $\tb \in \mathbb N_0$, the digraph $\Gamma_\tb$ is the Hasse diagram of poset $(\mathcal{O}_\tb, \ge)$, where a pair of feasible outcomes $o_1$ and $o_2$ satisfies $o_1 \ge o_2$ if and only if there is a directed path in $\Gamma_\tb$ from $o_2$ to $o_1$.
\end{thm}

\begin{proof}
    This is direct by construction and the definition of short form feasible outcome.
\end{proof}

Next, we prove that given a $\tb \in \mathbb N_0$, $(\mathcal{O}_\tb, \ge)$ is a lattice. 
 \begin{thm}\label{thm:lattice} For any $\tb \in \mathbb {N}_0$, consider any short forms $(a, b), (a', b')\in\mathcal{O}_\tb$ with $a\le a'$. 
Then the poset $(\mathcal{O}_\tb, \ge)$ is a lattice with join $``\vee"$ and meet $``\wedge"$ given by 
\begin{align*}
    (a,b)\vee (a', b') &= \begin{cases}
			(a,b) & \text{if } b\le b' \\
			(a,b') & \text{otherwise } 
		\end{cases}\\
  (a,b)\wedge (a', b') &= \begin{cases}
			(a',b') & \text{if } b\le b' \\
			(a',b) & \text{otherwise}.
		\end{cases}
\end{align*}
 \end{thm}
 \begin{proof}
     
    Consider the case when $a\le a'$ and $b \le b'$. In this case $(a, b) $ and $(a', b')$ are comparable. Thus $(a,b)\vee (a', b') = (a,b)$ and $(a,b)\wedge (a', b') = (a', b')$. 
    
    Now we look into the other case. By Marker worth and the assumption $a\le a'$, observe that $b>b'$ implies $b'<b\le a+1 \le a'+1$. Hence  $(a,b'), (a',b) \in \mathcal{O}_\tb$.
    
In the case when $a=a'$ and $b>b'$, the short forms $(a, b) $ and $(a', b')$ are comparable. Thus $(a,b)\vee (a', b') = (a',b') = (a,b')$ and $(a,b)\wedge (a', b') = (a, b)=(a', b)$.

    Next assume that $a<a'$ and $b>b'$. In this case the short forms $(a, b) $ and $(a', b')$ are incomparable. Let $(a'', b'') \in \mathcal{O}_\tb$ be the upper bound for both short forms. Thus, $a \ge a'', b \ge b''$ and $a' \ge a'', b' \ge b''$. By the assumption $a< a'$ we have $a'' \le a <a'$ and $b>b' \ge b''$. Hence $(a,b)\vee (a', b') = (a,b')$. By similar analysis we get $(a,b)\wedge (a', b') = (a', b)$.
 \end{proof}
 
\section{Future directions}\label{sec:future}
This work has produced a complete solution for the structure of the outcome classes of bidding combinatorial games that generalize alternating normal play. The most natural application of these outcomes is to play a disjunctive sum of games. Given a finite number of game components, a current player moves in precisely one of the game components. A sum of two games $G$ and $H$ is written $G+H$. The outcome alone does not suffice to understand such compositions of games, but one can define a more comprehensive partial order by letting $G\ge H$ if, for all games $X$, $o(G+X)\ge o(H+X)$. This inequality defines equivalence classes of games. In our follow up paper \cite{kant2022constructive}, we find a constructive solution for game comparison given a certain proviso. We refer the reader to that paper for further open problems and future directions.\\

\noindent{\bf Acknowledgement:} Theorem~\ref{thm:impartial} was discovered at the ``Chocolate Caf\'e'' between sessions at the CGTC3 in Lisbon 2019, and it inspired us to take a closer look at the proposed class of bidding combinatorial games.

\bibliographystyle{plain}
\bibliography{main1}

\end{document}